\documentclass[prl, twocolumn, amsmath, showpacs]{revtex4}
\usepackage{graphicx}

\begin{document}
\title{Universal optimal transmission of light through disordered materials}
\author{I.M. Vellekoop and A. P. Mosk}
\affiliation{Complex Photonic Systems, Faculty of Science and Technology, and MESA$^+$ Institute for
Nanotechnology, University of Twente, P. O. Box 217, 7500 AE Enschede, The Netherlands}

\date{\today}

\newcommand{\avg}[1]{\left\langle#1\right\rangle}
\newcommand{\bucket}{C_{4,2}}
\newcommand{\mum}[1]{#1\,\mu\text{m}}
\newcommand{\mm}[1]{#1\,\text{mm}}
\newcommand{\nm}[1]{#1\,\text{nm}}
\newcommand{\Ttot}{T_\text{tot}}               
\newcommand{\optimTtot}{T^{\text{opt,}\beta}_\text{tot}}
\newcommand{\actTtot}{T^\text{act}_\text{tot}} 
\newcommand{\actEbeta}{E^\text{act}_\beta}     
\newcommand{\optimEin}{E^{\text{opt,}\beta}_a} 
\newcommand{\actEin}{E^{\text{act,}\beta}_a}   
\newcommand{\Tcontrol}{T_\text{c}}
\newcommand{\Tr}{\text{Tr}\;}
\newcommand{\Treverse}{T_\beta}
\newcommand{\Tau}{\mathcal{T}}

\begin{abstract}
\noindent We experimentally demonstrate increased diffuse
transmission of light through strongly scattering materials.
Wavefront shaping is used to selectively couple light to the open
transport eigenchannels, specific solutions of Maxwell's equations which the sample transmits fully,
resulting in an increase of
up to 44\% in the total angle integrated transmission
compared to the case where plane waves are incident.
The results
for each of several hundreds of experimental runs are in excellent
quantitative agreement with random matrix theory.
From our measurements we conclude that with perfectly shaped wavefronts
the transmission of a disordered sample tends to a universal value
of $2/3$, regardless of the thickness.
\end{abstract}\maketitle

\noindent The transport of waves in strongly scattering media is,
usually, well described by diffusion theory. However, the diffusion
equation does not take into account interference. Interference gives
rise to fundamental effects such as enhanced
backscattering\cite{Albada1985, Wolf1985} and Anderson localization
of light\cite{Wiersma1997, Storzer2006, Schwartz2007}.
Such interference effects may be observed with incident light in a single freely propagating mode
(scattering channel), such as a plane incident wave.
In 1984, Dorokhov predicted a striking multichannel interference effect. Using random matrix theory (RMT),
he showed that the transmission through a diffusive material is the result of a small number of open
eigenchannels with a transmission coefficient of close to one.\cite{Dorokhov1984} Each
eigenchannel\footnote{Eigenchannels are specific to a single realization of disorder; they are a good
basis to describe propagation of waves inside the sample. We use the term `free modes' for an orthogonal
basis that describes free propagating waves outside the sample.} corresponds to a specific linear
combination of multiple free modes. Waves coupled to an open eigenchannel will be fully transmitted
through a disordered sample, even if the sample is optically thick. 
In contrast, an
 incident plane wave couples mainly to
closed eigenchannels (with a transmission coefficient of close to
zero) and, consequently, most of its power is diffusely reflected. This result
was originally
obtained
for electron transport in a wire. Later, it
was generalized to a slab geometry\cite{Nazarov1994} and to optical
systems\cite{Pendry1990,Beenakker1997}.

Universal conductance fluctuations, which are
random fluctuations in the coupling of waves to open and closed
eigenchannels, have been observed in experiments
\cite{Washburn1988, Boer1994, Scheffold1998, Hemmady2005}, however, no controlled coupling to open
eigenchannels has been reported so far. Here we experimentally
demonstrate the injection of light into the open eigenchannels of a
strongly disordered sample,
resulting in a large increase of the diffusely transmitted intensity. We accomplish this by shaping the
wavefront of the incident light, which amounts to individually controlling the phases of multiple incoming
free modes.

The transmission amplitudes of waves through a mesoscopic medium
are
described
by a transmission matrix $t$. The field in the outgoing free modes at the back of the sample is given by
\begin{equation}
E_b = \sum_a^N t_{ba} E_a,\label{eq:t-def}
\end{equation}
where indices $a$ and $b$ label incident and transmitted free modes respectively. $N$ is the total number
of incident free modes. Any complete orthogonal set of modes may be chosen; in mesoscopic physics it is
usual to choose the transversal modes of a perfect waveguide as a basis, one may also work with an
overcomplete basis of incident angles or diffraction limited spots on the sample surface. Transmission
eigenchannels are defined for each individual sample by decomposing the transmission matrix as $t=U\Tau
V^T$. Here $U$ and $V$ are unitary matrices that effect the basis transformations between free modes
outside the sample and eigenchannels inside. $\Tau$ is a real diagonal matrix containing eigenchannel
transmission coefficients.


\begin{figure}
\centering
  \includegraphics[width=8.6cm]{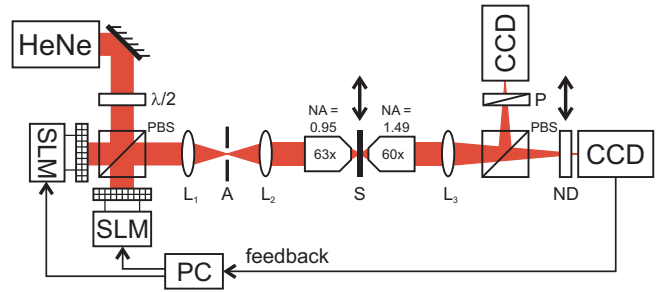}\\
  \caption{Experimental setup.
  HeNe, expanded $\nm{632.8}$ HeNe laser; $\lambda/2$, half waveplate; PBS,
  polarizing beam splitter cube; SLM, spatial light modulator. A, iris diaphragm; 63x, microscope
  objective; 60x, oil immersion microscope objective; S, sample; P, polarizer; ND, neutral density filter;
  L$_1$, L$_2$, L$_3$, lenses with focal length of respectively $\mm{250}$, $\mm{200}$ and $\mm{600}$.
  ND and S are translated by computer controlled stages.
}\label{fig:setup-dorokhov}
\end{figure}


We use a spatial light modulator (see Fig.~\ref{fig:setup-dorokhov}) to shape the wavefront of the
incident light. The computer program that controls the wavefront shaper optimizes the intensity of a
small target spot at the back surface of the sample using the measured intensity in the target as feedback \cite{Vellekoop2007}. This
wavefront shaping method creates a high intensity focus through the sample, which by itself is not a result of the bimodal
distribution of the eigenchannel transmission coefficients. Importantly, the disordered background speckle
\emph{around} the focus turns out to be a very sensitive probe of the distribution of the elements of
$\Tau$.

For example, for the limiting distribution were all eigenchannels are completely open $\Tau=0$ or
completely closed $\Tau=1$ (the so called maximal fluctuations \cite{Pendry1990}), only open eigenchannels
will contribute to the intensity in the target.  Therefore, the optimization algorithm will select only
linear combinations corresponding to open eigenchannels. As a result, the diffuse transmission will
increase so that the ideally shaped wavefront will have a total transmission of unity. The distribution
predicted by Dorokhov\cite{Dorokhov1984} is close to this limiting case.

We now proceed to introduce a quantitative measure of the control we exert over a shaped wavefront. Our
algorithm maximizes the intensity in a diffraction limited spot, which is exactly one of the transmitted free modes. We label this special target mode
with the index $\beta$. The ideally shaped incident wavefront $\optimEin$ for maximizing the intensity in
$\beta$ is given by \cite{PRL-EPAPS},
\begin{equation}
\optimEin = \left(\Treverse\right)^{-1/2} t^*_{\beta a}, \qquad \Treverse\equiv \sum_a^N |t_{\beta
a}|^2\label{eq:optimal-field},
\end{equation}
where $\Treverse$ normalizes the total incident power. Our optimization algorithm proceeds as follows: the
matrix elements $t_{\beta a}$ are measured up to a constant prefactor by cycling the phase of the light in
the incident mode $a$ while observing the intensity in target mode $\beta$ \cite{Vellekoop2007}. After $N$
phases have been measured, the optimized incident wavefront is constructed according to
Eq.(\ref{eq:optimal-field}). This optimized wavefront couples to a superposition of eigenchannels, mostly
to channels with high transmission eigenvalues.

In any experiment, the resolution and the spatial extent of the
generated field are finite. Therefore, it will never be possible to
exactly construct the wavefront described by
Eq.~\eqref{eq:optimal-field}. To quantify how well the actual
incident field $\actEin$ matches the optimal incident field
$\optimEin$, we introduce the overlap coefficient $\gamma$ as
\begin{equation}
\gamma \equiv \sum_a^N \left( \optimEin \right)^* \actEin\label{eq:gamma-definition}.
\end{equation}
The degree of intensity control is
 $|\gamma|^2$. We can now write any incident wavefront as a
linear superposition of the perfect wavefront and an error term
\begin{equation}
\actEin = \gamma \optimEin + \sqrt{1-|\gamma|^2} \Delta E_a,\label{eq:gamma-decomposition}
\end{equation}
where the error term $\Delta E_a$ is normalized. For ideal control over the incident wavefront
$|\gamma|^2=1$. In earlier
experiments\cite{Vellekoop2007} the degree of control was relatively low ($|\gamma|^2\ll 0.1$), and total
transmission did not increase measurably. In this Letter, we discuss experiments at much higher values of
$|\gamma|^2$, up to 0.33.

The experimental apparatus (see Fig.~\ref{fig:setup-dorokhov}) is designed to approach the optimal
wavefront as closely as possible, by controlling the largest possible fraction of the incident free modes.
An expanded beam from a $\nm{632.8}$ HeNe laser is rotated to a $45^\circ$ linear polarization by a half
waveplate and impinges on a polarizing beam splitter cube. Horizontally and vertically polarized beams are
modulated with separate reflective liquid crystal displays (Holoeye LC-R 2500) and then recombined, to
provide control over modes with both polarizations. We used a 4-pixel macropixel modulation
method\cite{Putten2008} to control the phase of the light without
residual amplitude modulation. The modulator is divided into 3816 independently programmable segments. A
computer programs the modulators
using feedback from a camera as discussed below. A sequential optimization algorithm\cite{Vellekoop2007}
was used to optimize the wavefront\cite{PRL-EPAPS}. A high numerical aperture objective ($\text{NA}=0.95$,
Zeiss Achroplan $63\times$) projects the shaped wavefront onto the sample.

Each sample consists of a layer of spray-painted ZnO particles on a standard glass microscope cover slip.
The particles have an average diameter of $\nm{200}$, which makes them strongly scattering for visible
light. The mean free path was determined by measuring the total transmission and equals
$\mum{0.85\pm0.15}$ at a wavelength of $\nm{632.8}$. We used samples with thicknesses of $\mum{5.7}$ and
$\mum{11.3}$. The samples were positioned in the focal plane of the microscope objective to minimize the
size of the diffuse spot, and thereby the number of contributing modes. The number of such contributing
free modes was estimated from the intensity profile of the transmitted light\cite{Boer1994} to be
$5.5\cdot 10^3$ and $1.0\cdot 10^4$ modes, for the thin and the thick samples, respectively.  The samples
were mounted on a motorized stage to translate them in the focal plane.

A high NA oil-immersion objective (Nikon TIRF $60\times$/NA$=1.49$) collects the transmitted light.
The transmitted light is split into horizontal and vertical polarizations by a beam splitter cube. A
second polarizer improves the extinction ratio for reflected light. The magnification of the detection
system is $225\times$, enough to well resolve individual speckles. A camera measures the power of the
horizontally polarized light in a disc with a diameter of $\mum{0.11}$ at the sample,
which is smaller than a single speckle, to provide feedback for the optimization algorithm.

After optimization, a calibrated neutral density filter with a transmission of $1.4\cdot10^{-3}$ is placed
in front of the camera to measure the high intensity in the target. A second camera images the intensity
of the vertically polarized light.

\begin{figure}
\centering
\includegraphics[width=8.6cm]{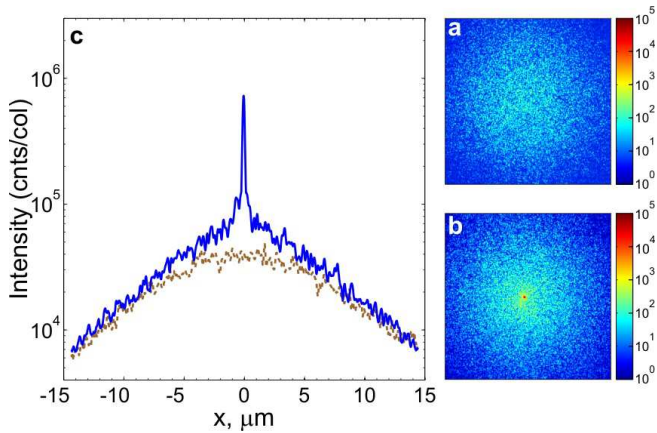}
  \caption{Intensity distribution of horizontally polarized light in a
  $\mum{15}\times\mum{15}$ square area at the back of the sample.
  (a) For a non-optimized incident wavefront. (b) For an optimized wavefront.
  (c) Intensity summed in the y-direction to average over speckle.
  Dashed curve, transmission of non-optimized wavefront;
  Solid curve, transmission of optimized wavefront.}\label{fig:results-before-after-dorokhov}
\end{figure}


Optimizing the incident wavefront caused the intensity in the target to increase dramatically. In
Fig.~\ref{fig:results-before-after-dorokhov} we plot the transmitted intensity through a
$\mum{11.3}$-thick sample for a non-optimized wavefront and for the optimized wavefront\cite{PRL-EPAPS}.
Before optimization, the transmitted intensity forms a diffuse spot on the back surface of the sample.
After optimization, a strong peak emerges in the target focus. The intensity increase in the center of the
target was a factor of $746\pm28$. After optimization, 2.3\% of the incident power is transmitted into the
target focus.

More importantly, the intensity in an area with a radius of approximately $\mum{5}$ around the target also
increased, even though the algorithm did not use this intensity as feedback. This observation indicates
that we have redistributed the incident light from closed eigenchannels to open eigenchannels. As a result
of optimizing a single target point, the total angle-integrated transmission increased from $0.23$ to
$0.31$. This change amounts to a relative increase of $35\%$.

For a quantitative analysis we need to know the degree of control $|\gamma|^2$. Factors like measurement
noise and thermal drift result in a different degree of control for each single run of the experiment.
Fortunately, it is possible to measure $|\gamma|^2$ directly for each run by observing the intensity in
the target. Only the controlled fraction of the incident wavefront contributes to the intensity in the
target. The transmission to the target mode $\beta$ equals
\begin{align}
\left| \actEbeta \right|^2 &= \left| \sum_a^N t_{\beta a} \actEin \right|^2\\
&= |\gamma|^2 \left| \sum_a^N t_{\beta a} \optimEin \right|^2
\end{align}
where we used Eq. \eqref{eq:gamma-decomposition} and the fact that the error term is orthogonal to the
ideal wavefront. By substituting Eq.~\eqref{eq:optimal-field} we obtain
\begin{equation}
\left| \actEbeta \right|^2 = |\gamma^2|
\Treverse.\label{eq:intensity-optimized-spot}
\end{equation}
Equation~\eqref{eq:intensity-optimized-spot} allows us to obtain the degree of control by measuring the
intensity in the target focus $\left| \actEbeta \right|^2$ and $\Treverse$. In the experimental procedure
it is very impractical to measure $\Treverse$. Therefore, we approximate $\Treverse = \Ttot N/M$. Here,
$\Ttot$ is the ensemble averaged total transmission of an unoptimized wavefront, and $M$ is the number of
transmitted free modes. Since our samples are sandwiched between a glass substrate on one side and air on
the other side, the number of modes on the back of the sample is larger and $M=n^2 N$, with $n=1.52$ the
refractive index of the substrate. This approximation neglects the $C_2$ fluctuations\cite{Berkovits1994}
in the total transmission, which are in the order of 2\% for our samples.

For the experimental run that is shown in Fig.~\ref{fig:results-before-after-dorokhov}, we find a degree
of control of $|\gamma|^2=0.23$. This means that the incident field is a linear superposition of the
perfectly shaped wavefront (carrying $23\%$ of the incident power) and a noise term (carrying the rest
of the power). The total, angle-integrated transmission $\actTtot$ contains contributions both from
the perfectly shaped wavefront and from the noise term,
\begin{equation}
\actTtot = \Tcontrol + \left(1-|\gamma|^2\right)\Ttot,\label{eq:total-transmission-optimized}
\end{equation}
where $\Tcontrol$ is the part of the transmission resulting from the perfectly shaped fraction of the
incident wavefront. By substituting Eq.~\eqref{eq:optimal-field} into Eq.~\eqref{eq:t-def} and summing the
power in all transmitted free modes, we find
\begin{equation}
\Tcontrol = |\gamma|^2 \sum_b^M \frac{1}{\Treverse} \left| \sum_a^N t_{ba} t_{\beta a}^* \right|^2 \equiv
|\gamma|^2 \bucket.\label{eq:bucket-def}
\end{equation}
We evaluate $\bucket$ theoretically by averaging over all possible target modes $\beta$. We assume that
$\bucket$ is self averaging, which is verified by our experiment. Neglecting small correlation terms between numerator and denominator we find
\begin{equation}
\bucket = \frac{1}{\avg{\Treverse}_\beta}\avg{\sum_b^M \left| \sum_a^N t_{ba} t_{\beta a}^*
\right|^2}_\beta = \frac{\Tr t^\dag t t^\dag t}{\Tr t t^\dag}\label{eq:bucket-avg}.
\end{equation}
 From
Eq.~\eqref{eq:bucket-avg}, it becomes clear that $\bucket$ is a measure for the width of the distribution
of the transmission eigenvalues\cite{Pendry1990}. By measuring the total transmission after optimizing the
incident wavefront, we have direct experimental access to this value for each single sample. Since we
measured $|\gamma|^2$ separately, we can use Eqs.~\eqref{eq:total-transmission-optimized} and
\eqref{eq:bucket-def} to obtain $\bucket$ from a single, non-ideal experimental run. In the particular run
in Fig.~\ref{fig:results-before-after-dorokhov}, we find $\bucket=0.62$.

The \emph{ensemble averaged} value for $\bucket$ was derived using RMT. RMT\cite{Mello1988} predicts
$\avg{\bucket}=2/3$ for a non-absorbing system far away from the localization transition, regardless of
the original transmission coefficient of the system. For a single realization of disorder, we found
$\bucket=0.62$. To investigate the universality of this result and to compare the measured values with
RMT, we performed automated sequences of measurements. The sample was translated between the runs to
obtain different realizations of disorder.

\enlargethispage{\baselineskip} 

The intensity increase was different for each of the optimizations in the sequence. The increase in the
integrated transmission varied from a few percent to a maximum of 44\%. These variations are the result of
drift due to varying environmental conditions. While drift is undesirable, it gives us a wide range of
$|\gamma|^2$ to investigate. The degree of control $|\gamma|^2$ was determined for each of the runs using
Eq.~\eqref{eq:intensity-optimized-spot}. Then, using Eq.~\eqref{eq:total-transmission-optimized}, we
isolated $\Tcontrol$.


\begin{figure}
\centering
  \includegraphics[width=8.6cm]{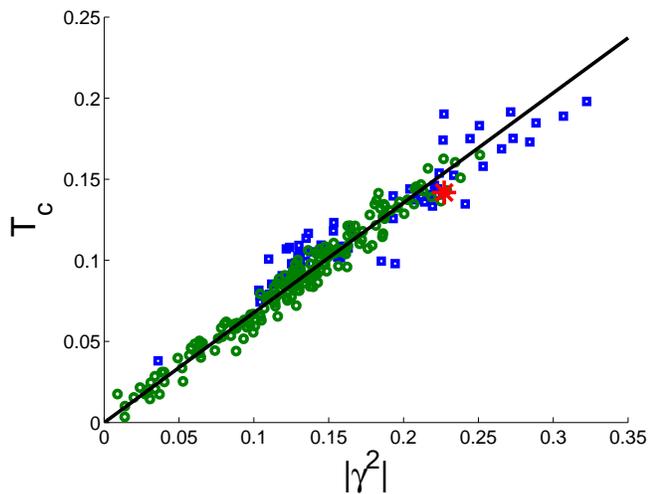}\\
  \caption{Total transmission
  of the controlled fraction of the wavefront.
  Data from two different samples
  collapses to a universal curve. There are no adjustable parameters in the data processing or the theory.
  Circles, results for $\mum{11.3}$-thick sample;
  squares, results for $\mum{5.7}$-thick sample;
  star, data point corresponding to the experimental run that was shown in detail in
  Fig.~\ref{fig:results-before-after-dorokhov};
  solid line, linear regression with a slope of 0.68;
  dotted line, slope of $2/3$ as predicted by RMT.}\label{fig:results-corrected-transmission}
\end{figure}

In Fig.~\ref{fig:results-corrected-transmission} we plotted $\Tcontrol$ versus $|\gamma|^2$ for hundreds
of measurements. To determine the effect of the sample thickness, two different samples were used: one
with a thickness of $\mum{5.7\pm0.5}$, and one that is approximately twice as thick ($\mum{11.3\pm0.5}$).
All data points collapse to a single line. Our data shows that the transmission coefficient of an ideally
shaped wavefront does not depend on the sample thickness. A linear regression gives
$\avg{\bucket}=0.68\pm0.07$ where the uncertainty follows from a worst case estimate of the systematical
errors in the experiment. This value is in excellent agreement with RMT.


We have shown experimental evidence of controlled coupling of light into open transmission eigenchannels
in opaque, strongly
scattering materials.
The coupling to open eigenchannels was enhanced by using a wavefront shaping algorithm to optimize
transmission to a focus, and detected by measuring the angle-averaged transmission intensity, which showed
a relative increase of up to 44\%. We quantitatively compared the results for different samples and
different realizations of disorder. All results showed a universal behavior that is in excellent
quantitative agreement with random matrix theory. Our results demonstrate that RMT of wave transport can
successfully be applied to open systems and single realizations of disorder. This conclusion is relevant
for the propagation of electromagnetic waves, matter waves and sound in open, strongly scattering
environments.

We thank Ad Lagendijk and Willem Vos for valuable advice, and Elbert van Putten and Frerik van Beijnum
for discussions. This work is part of the research program of the ``Stichting voor Fundamenteel Onderzoek
der Materie (FOM)", which is financially supported by the ``Nederlandse Organisatie voor Wetenschappelijk
Onderzoek" (NWO). A.~P.~Mosk is supported by a VIDI grant from NWO.

\bibliography{../../bibliography}

\begin{thebibliography}{18}
\expandafter\ifx\csname natexlab\endcsname\relax\def\natexlab#1{#1}\fi
\expandafter\ifx\csname bibnamefont\endcsname\relax
  \def\bibnamefont#1{#1}\fi
\expandafter\ifx\csname bibfnamefont\endcsname\relax
  \def\bibfnamefont#1{#1}\fi
\expandafter\ifx\csname citenamefont\endcsname\relax
  \def\citenamefont#1{#1}\fi
\expandafter\ifx\csname url\endcsname\relax
  \def\url#1{\texttt{#1}}\fi
\expandafter\ifx\csname urlprefix\endcsname\relax\def\urlprefix{URL }\fi
\providecommand{\bibinfo}[2]{#2}
\providecommand{\eprint}[2][]{\url{#2}}

\bibitem[{\citenamefont{van Albada and Lagendijk}(1985)}]{Albada1985}
\bibinfo{author}{\bibfnamefont{M.~P.} \bibnamefont{van Albada}}
  \bibnamefont{and}
  \bibinfo{author}{\bibfnamefont{A.}~\bibnamefont{Lagendijk}},
  \bibinfo{journal}{Phys. Rev. Lett.} \textbf{\bibinfo{volume}{55}},
  \bibinfo{pages}{2692} (\bibinfo{year}{1985}).

\bibitem[{\citenamefont{Wolf and Maret}(1985)}]{Wolf1985}
\bibinfo{author}{\bibfnamefont{P.~E.} \bibnamefont{Wolf}} \bibnamefont{and}
  \bibinfo{author}{\bibfnamefont{G.}~\bibnamefont{Maret}},
  \bibinfo{journal}{Phys. Rev. Lett.} \textbf{\bibinfo{volume}{55}},
  \bibinfo{pages}{2696} (\bibinfo{year}{1985}).

\bibitem[{\citenamefont{Wiersma et~al.}(1997)\citenamefont{Wiersma, Bartolini,
  Lagendijk, and Righini}}]{Wiersma1997}
\bibinfo{author}{\bibfnamefont{D.~S.} \bibnamefont{Wiersma}},
  \bibinfo{author}{\bibfnamefont{P.}~\bibnamefont{Bartolini}},
  \bibinfo{author}{\bibfnamefont{A.}~\bibnamefont{Lagendijk}},
  \bibnamefont{and} \bibinfo{author}{\bibfnamefont{R.}~\bibnamefont{Righini}},
  \bibinfo{journal}{Nature} \textbf{\bibinfo{volume}{390}},
  \bibinfo{pages}{671} (\bibinfo{year}{1997}).

\bibitem[{\citenamefont{St{\"o}rzer et~al.}(2006)\citenamefont{St{\"o}rzer,
  Gross, Aegerter, and Maret}}]{Storzer2006}
\bibinfo{author}{\bibfnamefont{M.}~\bibnamefont{St{\"o}rzer}},
  \bibinfo{author}{\bibfnamefont{P.}~\bibnamefont{Gross}},
  \bibinfo{author}{\bibfnamefont{C.~M.} \bibnamefont{Aegerter}},
  \bibnamefont{and} \bibinfo{author}{\bibfnamefont{G.}~\bibnamefont{Maret}},
  \bibinfo{journal}{Phys. Rev. Lett.} \textbf{\bibinfo{volume}{96}},
  \bibinfo{pages}{063904} (\bibinfo{year}{2006}).

\bibitem[{\citenamefont{Schwartz et~al.}(2007)\citenamefont{Schwartz, Bartal,
  Fishman, and Segev}}]{Schwartz2007}
\bibinfo{author}{\bibfnamefont{T.}~\bibnamefont{Schwartz}},
  \bibinfo{author}{\bibfnamefont{G.}~\bibnamefont{Bartal}},
  \bibinfo{author}{\bibfnamefont{S.}~\bibnamefont{Fishman}}, \bibnamefont{and}
  \bibinfo{author}{\bibfnamefont{M.}~\bibnamefont{Segev}},
  \bibinfo{journal}{Nature} \textbf{\bibinfo{volume}{446}}, \bibinfo{pages}{52}
  (\bibinfo{year}{2007}).

\bibitem[{\citenamefont{Dorokhov}(1984)}]{Dorokhov1984}
\bibinfo{author}{\bibfnamefont{O.~N.} \bibnamefont{Dorokhov}},
  \bibinfo{journal}{Sol. St. Commun.} \textbf{\bibinfo{volume}{51}},
  \bibinfo{pages}{381} (\bibinfo{year}{1984}).

\bibitem[{\citenamefont{Nazarov}(1994)}]{Nazarov1994}
\bibinfo{author}{\bibfnamefont{Y.~V.} \bibnamefont{Nazarov}},
  \bibinfo{journal}{Phys. Rev. Lett.} \textbf{\bibinfo{volume}{73}},
  \bibinfo{pages}{134} (\bibinfo{year}{1994}).

\bibitem[{\citenamefont{Pendry et~al.}(1990)\citenamefont{Pendry, MacKinnon,
  and Pretre}}]{Pendry1990}
\bibinfo{author}{\bibfnamefont{J.~B.} \bibnamefont{Pendry}},
  \bibinfo{author}{\bibfnamefont{A.}~\bibnamefont{MacKinnon}},
  \bibnamefont{and} \bibinfo{author}{\bibfnamefont{A.~B.}
  \bibnamefont{Pretre}}, \bibinfo{journal}{Physica A}
  \textbf{\bibinfo{volume}{168}}, \bibinfo{pages}{400} (\bibinfo{year}{1990}).

\bibitem[{\citenamefont{Beenakker}(1997)}]{Beenakker1997}
\bibinfo{author}{\bibfnamefont{C.~W.~J.} \bibnamefont{Beenakker}},
  \bibinfo{journal}{Rev. Mod. Phys.} \textbf{\bibinfo{volume}{69}},
  \bibinfo{pages}{731} (\bibinfo{year}{1997}).

\bibitem[{\citenamefont{Washburn}(1988)}]{Washburn1988}
\bibinfo{author}{\bibfnamefont{S.}~\bibnamefont{Washburn}},
  \bibinfo{journal}{J. Res. Develop.} \textbf{\bibinfo{volume}{32}},
  \bibinfo{pages}{335} (\bibinfo{year}{1988}).

\bibitem[{\citenamefont{de~Boer et~al.}(1994)\citenamefont{de~Boer, van Rossum,
  van Albada, Nieuwenhuizen, and Lagendijk}}]{Boer1994}
\bibinfo{author}{\bibfnamefont{J.~F.} \bibnamefont{de~Boer}},
  \bibinfo{author}{\bibfnamefont{M.~C.~W.} \bibnamefont{van Rossum}},
  \bibinfo{author}{\bibfnamefont{M.~P.} \bibnamefont{van Albada}},
  \bibinfo{author}{\bibfnamefont{T.~M.} \bibnamefont{Nieuwenhuizen}},
  \bibnamefont{and}
  \bibinfo{author}{\bibfnamefont{A.}~\bibnamefont{Lagendijk}},
  \bibinfo{journal}{Phys. Rev. Lett.} \textbf{\bibinfo{volume}{73}},
  \bibinfo{pages}{2567} (\bibinfo{year}{1994}).

\bibitem[{\citenamefont{Scheffold and Maret}(1998)}]{Scheffold1998}
\bibinfo{author}{\bibfnamefont{F.}~\bibnamefont{Scheffold}} \bibnamefont{and}
  \bibinfo{author}{\bibfnamefont{G.}~\bibnamefont{Maret}},
  \bibinfo{journal}{Phys. Rev. Lett.} \textbf{\bibinfo{volume}{81}},
  \bibinfo{pages}{5800} (\bibinfo{year}{1998}).

\bibitem[{\citenamefont{Hemmady et~al.}(2005)\citenamefont{Hemmady, Zheng, Ott,
  Antonsen, and Anlage}}]{Hemmady2005}
\bibinfo{author}{\bibfnamefont{S.}~\bibnamefont{Hemmady}},
  \bibinfo{author}{\bibfnamefont{X.}~\bibnamefont{Zheng}},
  \bibinfo{author}{\bibfnamefont{E.}~\bibnamefont{Ott}},
  \bibinfo{author}{\bibfnamefont{T.~M.} \bibnamefont{Antonsen}},
  \bibnamefont{and} \bibinfo{author}{\bibfnamefont{S.~M.}
  \bibnamefont{Anlage}}, \bibinfo{journal}{Phys. Rev. Lett.}
  \textbf{\bibinfo{volume}{94}}, \bibinfo{pages}{014102}
  (\bibinfo{year}{2005}).

\bibitem[{\citenamefont{Vellekoop and Mosk}(2007)}]{Vellekoop2007}
\bibinfo{author}{\bibfnamefont{I.~M.} \bibnamefont{Vellekoop}}
  \bibnamefont{and} \bibinfo{author}{\bibfnamefont{A.~P.} \bibnamefont{Mosk}},
  \bibinfo{journal}{Opt. Lett.} \textbf{\bibinfo{volume}{32}},
  \bibinfo{pages}{2309} (\bibinfo{year}{2007}).

\bibitem[{PRL()}]{PRL-EPAPS}
\bibinfo{note}{See EPAPS Document No. for details on methods and materials
  used. For more information on EPAPS, see
  http://www.aip.org/pubservs/epaps.html.}

\bibitem[{\citenamefont{van Putten et~al.}(2008)\citenamefont{van Putten,
  Vellekoop, and Mosk}}]{Putten2008}
\bibinfo{author}{\bibfnamefont{E.~G.} \bibnamefont{van Putten}},
  \bibinfo{author}{\bibfnamefont{I.~M.} \bibnamefont{Vellekoop}},
  \bibnamefont{and} \bibinfo{author}{\bibfnamefont{A.~P.} \bibnamefont{Mosk}},
  \bibinfo{journal}{Appl. Opt.} \textbf{\bibinfo{volume}{47}},
  \bibinfo{pages}{2076} (\bibinfo{year}{2008}).

\bibitem[{\citenamefont{Berkovits and Feng}(1994)}]{Berkovits1994}
\bibinfo{author}{\bibfnamefont{R.}~\bibnamefont{Berkovits}} \bibnamefont{and}
  \bibinfo{author}{\bibfnamefont{S.}~\bibnamefont{Feng}},
  \bibinfo{journal}{Phys. Rep.} \textbf{\bibinfo{volume}{238}},
  \bibinfo{pages}{135} (\bibinfo{year}{1994}).

\bibitem[{\citenamefont{Mello et~al.}(1988)\citenamefont{Mello, Pereyra, and
  Kumar}}]{Mello1988}
\bibinfo{author}{\bibfnamefont{P.~A.} \bibnamefont{Mello}},
  \bibinfo{author}{\bibfnamefont{P.}~\bibnamefont{Pereyra}}, \bibnamefont{and}
  \bibinfo{author}{\bibfnamefont{N.}~\bibnamefont{Kumar}},
  \bibinfo{journal}{Ann. Physics} \textbf{\bibinfo{volume}{181}},
  \bibinfo{pages}{290} (\bibinfo{year}{1988}).

\end{thebibliography}
\bibliographystyle{apsrev}
\end{document}